# WEB RESOURCE FOR STORING COLLECTIVE EXPERIENCE

Olegs Verhodubs

oleg.verhodub@inbox.lv

**Abstract.** Experience is what makes our life more effective that is why it is necessary to share experience among people. The use of information technologies is the most technological way to work with experience, and the use of the Web is the best way for sharing it. This paper describes a web resource designed for storing, sharing and using experience that is obtained from different people in the Web. The main purpose of this paper is to present this web resource in order to evaluate the interest in such a web resource.

I. INTRODUCTION

There are a lot of fields of science, where their own facts and laws are identified during many centuries. Despite the laws in different fields of science are different, it can be concluded that some laws from one field of science manifest themselves in a completely another one. For example, the operating principles of centripetal and centrifugal force known from physics discover themselves in the development of the Web. Centripetal force is a force on an object directed to the center of a circular path that keeps the object on the path [1]. Centrifugal force is a tendency for an object to leave the circular path and fly off in a straight line [1]. The structure of data placement in the Web is not static or constant, and it is transforming according to centripetal and centrifugal forces, which act in the environment of data. Data in the Web either gravitate to one center, or run away from the single center. The center here is a web page, where or from where data is aspired. Nowadays the process of data gravitation to a single center is predominant. For example, all personal data is accumulated in social networks like Facebook [2]. In turn, professional data is accumulated in social networks like LinkedIn [3] or ResearchGate [4]. The process of data gravitation to a single center concerns another types of categorized data, too. For instance, the data in the format of video strive to be represented in YouTube [5]. As well, the data in the format of pictures strive to be represented in Instagram [6]. Pictures, videos or personal data are not all kinds of possible data. Knowledge in the form of IF..THEN rules is another kind of data, because data, information and knowledge in total represent something in common that is a single substance, which may be perceived and interpreted differently namely as data, information and knowledge depending on the purpose of the perception or interpretation [7]. There is no single center of knowledge (rules) they would strive to. That is there is no any web resouse, which would accumulate knowledge (rules) as YouTube do with videos. Although, if acting forces in the Web are determined correctly, the objective necessity sooner or later will cause such a web resource to appear. Knowledge are very important, because they do our life more effective and do us more competitive. It would be strange, if there would no attempts to utilize information technologies in the Web for work with knowledge. These attempts were made, but the results of these attempts could not be considered optimal. For example, YouTube has not only a lot of entertaining videos, but also plenty of training videos for different areas. It could be assumed that Youtube was designed for storing and watching entertaining videos precisely and then this web resource was found sufficient to store and watch more serious content. Fairly, the use of the functional for the purposes that are not envisaged in the design process testifies not only to the success of the idea, but it also testifies that there are no other possibilities to realize these purposes. That is it testifies to the need for something separate to realize purposes, which are unaccustomed for other systems or resources. Indeed, YouTube is also useful as educational resource, because this web resource has a lot of training videos, which contain plenty of knowledge. However the usefulness of YouTube is not absolute. For example, it is necessary to watch some useful video until the end to know something that would know much faster if knowledge would be represented in another way. That is why a separate Web resource is necessary to store collective knowledge and experience. Knowledge and experience can be stored in the form of IF..THEN rules. This is rather technological way to work with, because there are a lot of developed reasoners, which give opportunity to operate with rules in the form of IF..THEN.

Actually, the prototype of the web resource for storing collective knowledge and experience has already been developed. This prototype was available via the Web, however it has bounded functional. Nevertheless this prototype has a serious potential for development. In this regard, the purpose of this paper is to evaluate the interest to such a prototype. This is important to decide wheather it makes sense to deploy the potential of the prototype or not.

This paper is divided into several sections. The next section describes the architecture of the prototype for storing collective knowledge and experience. Section III presents the work order with the prototype and some possible directions of its improvement. The following section is dedicated to the one more way for ontology generation. The last section expounds conclusions of this work.

II. ARCHITECTURE OF THE PROTOTYPE

YouTube is not the only web resource, which is used to store knowledge from different users. Unlike YouTube that is designed for entertainment mostly, but is also used for educational purposes, there is a whole class of web resources that initially are designed for experience exchange. This class of Web resources is web forums. A web forum is a website or section of a website that allows visitors to communicate with each other by posting messages [8]. Web forums are available for all kinds of topics: software support, help for webmasters, and programming discussions. While lots of web forums focus on IT (Information Technology) topics, they are not limited to information technology. There are forums related to health, fitness, cars, houses, teaching, parenting, and thousands of other topics. Some forums are general, like a fitness forum, while others are more specific, such as a forum for yoga instructors [8]. Having the advantage of collective content update, what allow to accumulate a lot of knowledge related to a particular topic, a web forum at the same time has one big disadvantage, which is that it often takes a long time to look through all user messages to find the information you need in the Web forum. Furthermore, the messages in any web forum have no structure that complicates perception of their meaning. In fact, each message in a web forum can be an arbitrary set of words or even characters. That is why the value of user's message on a web forum solely depends on the user's prudence and responsibility. Ranking of web forum participants in accordance with their authority, which is based on the value of the messages largely removes the problem. The messages of the web forum participant with higher authority can be taken into account more readily than the messages of the web forum participant, who has lower authority. Of course, the value of a message of one web forum participant is estimated by other web forum participants. The more valued messages, the greater the authority of their author. This is rather useful mechanism that does not lose its effectiveness.

Thus, it is possible to propose functional requirements to the prototype of a web resource for storing collective knowledge. This prototype has to contain the following possibilities:

1) searching for stored knowledge;
2) adding of knowledge;
3) deleting of knowledge;
4) assessment of knowledge;
5) registering of users.

All of these tasks are available via the Web, because this gives an opportunity to be available for plenty of users in different places in the world. The task of user registering is a trivial task that is realized in many web applications. The task is performed using MySQL [9] database management system and PHP scripting language [10]. The tasks of adding, deleting and searching for knowledge are realized by means of Java servlet [11] and Apache Lucene technology [12]. This prototype is placed in Apache HTTP Server [13] with Apache Tomcat [14], which is necessary for storing servlets. Considering all the above, the architecture of the prototype of a web resource for storing collective knowledge is the following (Fig.1.):

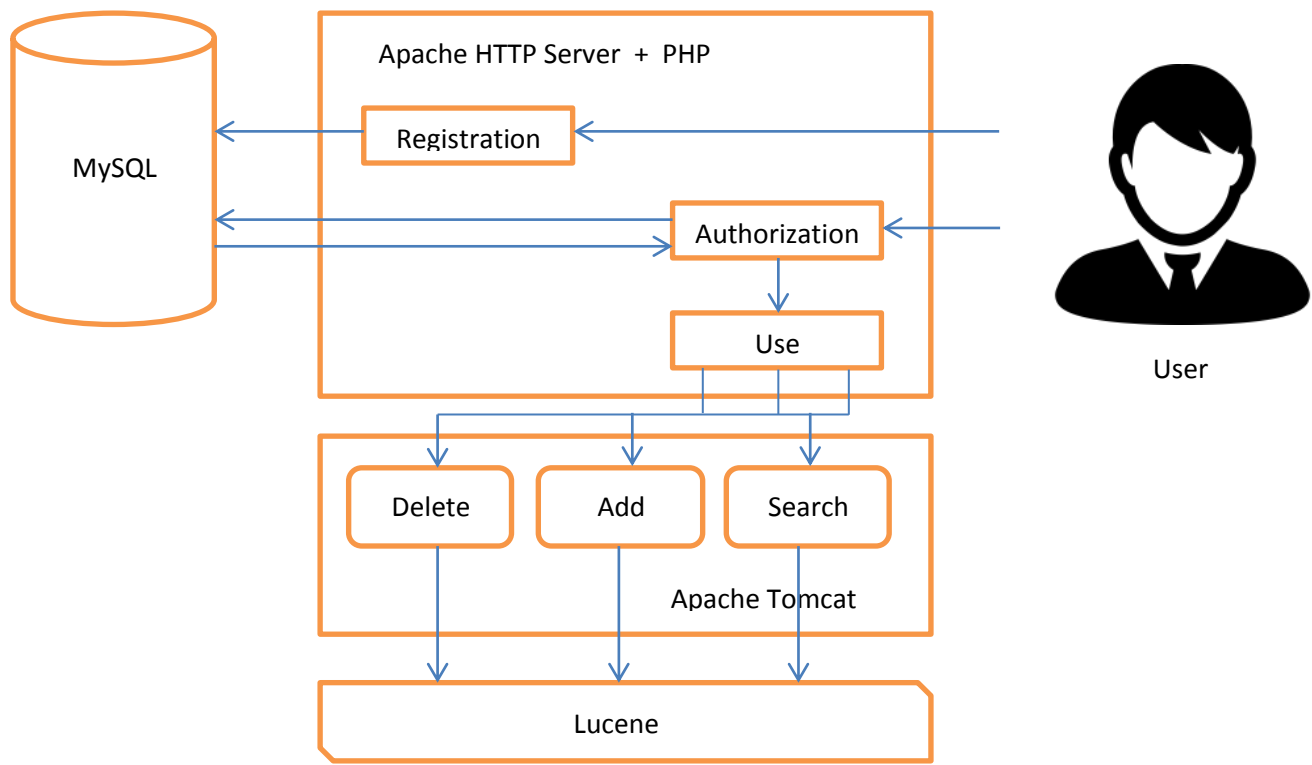

Fig. 1. Architecture of the prototype.

It is possible to exclude the PHP from the architecture, because PHP is necessary to access the MySQL database, but there is an opportunity to access the database by means of the Java programming language. PHP was chosen because the access the MySQL through PHP seemed faster than through the Java functions. This has to be tested professionally, but now this is not the first priority.

III. WORK WITH THE PROTOTYPE

The prototype is available at http://62.205.241.58. A domain name for this prototype is not purchased because currently the prototype is not intended for general use but for test use only. The architecture of the prototype consists of certain parts that determine the operating procedure with this prototype. Initially, the user has three ways to start working with the prototype. The first way is to register itself on the web resource. This means that the information about a user is saved in the MySQL database. The information is a name, e-mail and a password. Here everything is usual as in many other websites. This information is necessary to identify registered user from other users. The second way is to login the web resource and further to use the functional possibilities of the web resource. This way implied that the user has already been registered in the web resource. The functional possibilities of the web resource are a search for necessary knowledge and editing of personal knowledge in the web resource. Editing of personal knowledge is either adding new knowledge to the storage, or deleting the existing knowledge from this storage. Here knowledge are rules in the form of IF..THEN. It is necessary to add the IF and THEN parts to add the whole rule to a knowledge store. The IF and THEN parts are deleted from the knowledge store, when a rule is deleted. The third way is to use the search function of necessary knowledge at once. This way does not imply the login the website. It was possible to work in the web resource immediately as soon the website is loaded. Necessary concepts have to be inputted in the string of queries and the website outputs its answer and two possible reactions from the user. These possible reactions are presented as two buttons. The first button is "Yes" button, and the second one is "No" button. After the website outputted the answer, the user could press "Yes" button if he was satisfied with the answer or "No" button otherwise. For example, there are three rules in the storage of the web resource:

IF fly THEN bird

IF fly THEN plane

IF fly THEN rocket

Let us suppose that the user types a word "fly" in the string of the queries. So, the process of communication between the system and the user is the following:

System:> bird, isn't it?    [yes]   [**no**]

System:> plane, isn't it?   [**yes**]   [no]

System:> The result is plane

Here „no" button was pressed by the user in the first case, and „yes" button was pressed by the user in the second case. The prototype outputs the result „plane".

Certainly the example is primitive, but it is not difficult to develop the reasoning system. The development of the reasoning system can move in at least two directions. The first direction is to develop more complex reasoning that would take into consideration the interim facts, reasoned in the process of communication with the user. The second direction is to utilize the possibility of fuzzy reasoning. There are several ways how the possibility of fuzzy reasoning can be implemented in the web resource. However there is one more interesting possibility that can be realized by means of this website. This is a generation of ontology.

IV. GENERATION OF ONTOLOGY

The prototype is designed to accumulate and utilize knowledge gained from the users via the Web. The interest and participation of many people in the development of this website is its engine that keeps this website necessary for broad audience. Nevertheless broad audience and categorization of data entry, which consists of a separate premise and conclusion input, make it possible to extract an additional benefit that is ontology generation. Although the task of ontology generation is not a direct task of the website, this website is exactly the case, when side effects are just as important as the direct effects of the website for which it is designed. Strictly speaking ontology generation is not the side effect of realizing this website, because ontology is not generating by its own. So, the better name is side purpose, which nevertheless is conceived at the moment of designing.

The best resource for ontology generation is a raw text from the point of view of the richness and diversity of the human language. The richness of the human language is an advantage in terms of the amount of information that means the more the richness is the more information is. From the other hand the richness of the human language complicates retrieving of the information from the raw text. This means that something is necessary to reduce the complexity of retrieving of the information from the raw text. One of the few principles that we have is divide and rule or in other words division of the whole into parts. It is much easier to cope with the task in parts.

Any text consists of sentences that are units of meaning. It is difficult to divide this unit of meaning into parts without loss or distortion of meaning. In this regard division of information into rules that is IF and THEN parts can reduce the complexity of retrieving of information without loss of its meaning. The basis is the assumption that both IF and THEN parts presents self-sufficient in terms of meaning portion of information. Here self-sufficiency in terms of meaning is a semantic completeness, which does not need information from THEN part or information from other rules without fail. Consequently, having a lot of rules, which are inputted by users, it is possible to extract completed portions of meaning easily, because it is enough to get IF and THEN parts of these rules. Thus, IF and THEN parts is a set of statements. The order of these statements is not important in contrast to the order of IF and THEN parts in the rule. It is necessary to transform all statements from the set to ontology elements to generate the whole ontology (Fig.2.).

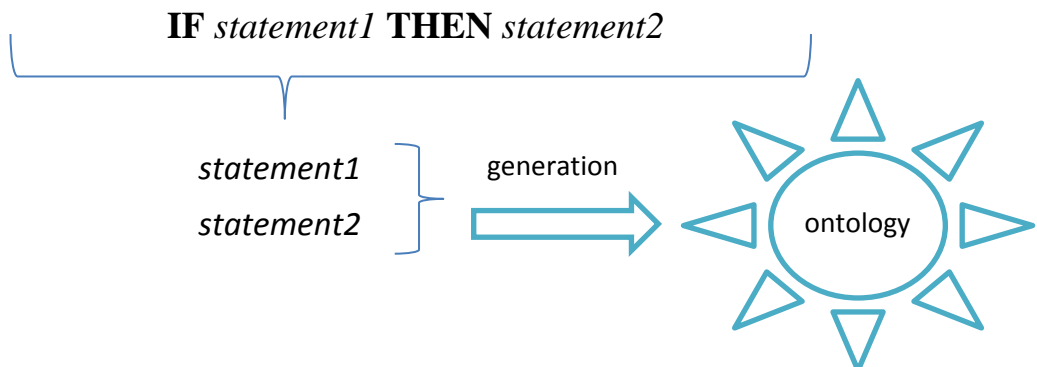

Fig.2. Ontology generation.

Each statement that is IF and THEN part has to be mapped into the OWL (Web Ontology Language) syntax. In terms of OWL each IF and THEN part has to generate classes, datatype and object properties. There are several ways to generate OWL classes, datatype and object properties from each statement. The simplest way is to detect parts of speech in each statement and generate the OWL element depending on this. There are several other ways to perform this task. The complexity is the structure of each statement can have complicated nature that is each statement can divide on several statements, which are connected by means of OR, AND, XOR connectors. In such a case, it is necessary to divide each IF or THEN statement on several statements as is shown below (Fig.3):

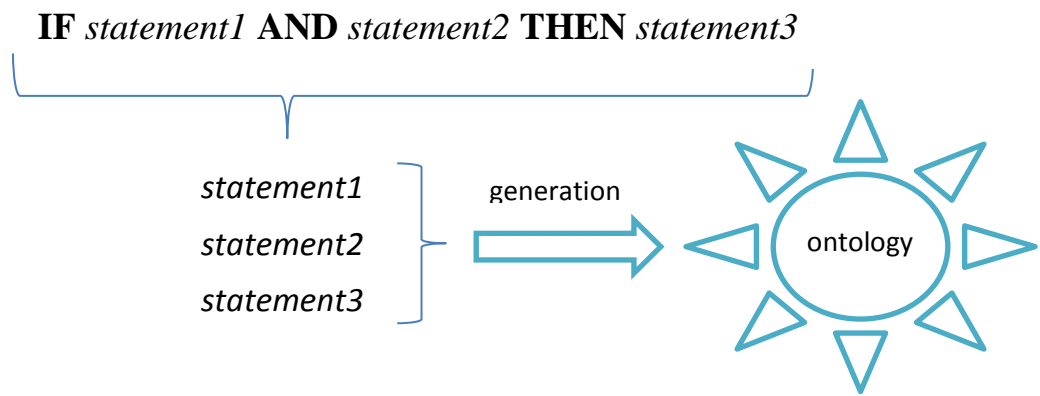

Fig.3. Ontology generation from complex rules.

It is difficult to imagine that every user will manually divide IF and THEN parts of the rule on the statements, which are strictly separated by means of AND, OR and XOR connectors. Most likely some users will not do this. Adding a rule some users will describe some situations or in other words states in IF and THEN parts, so this will require additional efforts to select the statements, which can be converted into the OWL syntax. In fact, this task is very similar to the task of ontology generation from raw text. The task is one of the most important in the Semantic Web and there are a lot of researchers, which go in for this task. However the task of rule generation instead of ontology generation from raw text seems more pragmatic from the point of knowledge use. This theme has already been considered in one of our previous research [7].

V. CONCLUSION

This paper describes the prototype of the web resource aimed to accumulate knowledge from users via the Web. The idea itself of accumulating knowledge in one place is not new. There are plenty of web forums in the Web, where people can share their thoughts, ideas, knowledge and even emotions. However technical realization of the existing web forums is rather archaic and it is about the same age as the Web. The presented prototype of the web resource shows the possible direction in the development of the web forums.

The advantage of the presented prototype for knowledge accumulation is that the prototype can be utilized to generate ontology. This is justified by the way of adding knowledge to the knowledge base of the prototype namely adding each unit of knowledge in parts that is condition and result separately. This way can have some difficulties in terms of ontology generation, but it seems these difficulties can be overcome in principle.

Developing of the presented prototype identified the need of raw text transformation to ontology. It is necessary to notice that if some algorithm of raw text transformation to ontology would suit completely, the presented prototype would not be needed at all. This is so, because ontology would be generated from raw text, but knowledge that are rules would be generated from ontology as is shown in previous papers [15][16]. It is much more promising to explore the transformation of raw text into rules directly. Such a possibility allows identifying the structure of any text from the point of view of internal laws of the phenomena, described in the text, but not from the point of view of parts and terms. It is possible to develop a universal expert system, which draws knowledge from the Web if it would be possible to generate rules from raw text.